\documentclass{article}
\usepackage{spconf,amsmath,graphicx,mathtools,amssymb,subfigure,float,hyperref,array,amsfonts,calrsfs,booktabs,multirow,siunitx,etoolbox,xcolor}

\usepackage{pifont}
\newcommand{\xmark}{\ding{55}}%

\title{AUDIO-VISUAL SPEECH INPAINTING WITH DEEP LEARNING}
\name{Giovanni Morrone$^1$ \qquad Daniel Michelsanti$^2$ \qquad Zheng-Hua Tan$^2$ \qquad Jesper Jensen$^{2,3}$}
\address{$^1$ University of Modena and Reggio Emilia, Department of Engineering "Enzo Ferrari", Italy \\
$^2$Aalborg University, Department of Electronic Systems, Denmark \\
$^3$Oticon A/S, Denmark
}
\begin{document}
%
\maketitle
\begin{abstract}
In this paper, we present a deep-learning-based framework for audio-visual speech inpainting, i.e.,\ the task of restoring the missing parts of an acoustic speech signal from reliable audio context and uncorrupted visual information. Recent work focuses solely on audio-only methods and generally aims at inpainting music signals, which show highly different structure than speech.
Instead, we inpaint speech signals with gaps ranging from 100 ms to 1600 ms to investigate the contribution that vision can provide for gaps of different duration. 
We also experiment with a multi-task learning approach where a phone recognition task is learned together with speech inpainting.
Results show that the performance of audio-only speech inpainting approaches degrades rapidly when gaps get large, while the proposed audio-visual approach is able to plausibly restore missing information. In addition, we show that multi-task learning is effective, although the largest contribution to performance comes from vision. 

\end{abstract}

\begin{keywords}
speech inpainting, audio-visual, deep learning, face-landmarks, multi-task learning
\end{keywords}
\section{Introduction}
\label{sec:intro}
In real life applications, audio signals are often corrupted by accidental distortions. Impulsive noises, clicks and even transmission errors might wipe out audio intervals. The process of restoring the lost information from the audio context is known as \emph{audio inpainting} \cite{adler_audio_2012}, and, when applied to speech signals, we refer to it as \emph{Speech Inpainting} (SI). Since human speech perception is multimodal, the use of visual information might be useful in restoring the missing parts of an acoustic speech signal. Visual information was successfully used in many speech-related tasks, such as speech recognition, speech enhancement, speech separation, etc. (cf. \cite{zhu2020deep, michelsanti2020overview} and references therein), but it has not been adopted for SI yet. In this paper, we address the problem of \emph{Audio-Visual Speech Inpainting} (AV-SI), i.e. the task of restoring the missing parts of an acoustic speech signal using audio context and visual information. 

The first audio inpainting works aimed at restoring short missing gaps in audio signals \cite{adler_audio_2012, godsill2002digital, smaragdis2009missing, wolfe2005interpolation}.
For inpainting long gaps, i.e.,\ hundreds of milliseconds, several solutions have been proposed. Bahat et al. \cite{bahat2015self} tried to fill missing gaps using pre-recorded speech examples from the same speaker and Perraudin et al.  \cite{perraudin_inpainting_2018} exploited self-similarity graphs within audio signals. However, the first approach required a different model for each speaker and the second one was less suitable for speech, since it could only inpaint stationary signals. Prablanc et al. \cite{prablanc_text-informed_2016} proposed a text-informed solution to inpaint missing speech combining speech synthesis and voice conversion models.

Several researchers attempted to solve audio inpainting using deep learning. In \cite{marafioti_context_2019}, a Convolutional Neural Network (CNN) model was used to inpaint missing audio from adjacent context. Other works exploited Generative Adversarial Networks (GANs) to generate sharper Time-Frequency (TF) representations \cite{ebner_audio_2020, marafioti_2020_gacela}. Recently, Zhou et al.\ \cite{zhou_vision-infused_2019} demonstrated that exploiting visual cues improved inpainting performance. However, these approaches only restored music signals, which usually have long term dependencies, unlike speech. Chang et al. \cite{chang_deep_2019} and Kegler et al.~\cite{Kegler2020} both tried to generate speech from masked signals with convolutional encoder-decoder architectures. They evaluated their systems on long gaps (about 500 ms), while in our work we aim at inpainting also extremely long segments (until 1600~ms), where additional information, like video, is essential to correctly restore speech signals.
A very recent work proposed a two-stage enhancement network where binary masking of a noisy speech spectrogram was followed by inpainting of time-frequency bins affected by severe noise~\cite{hao_masking_2020}.


In this paper, we propose a deep learning-based approach for speaker-independent SI where visual information is used together with the audio context to improve restoration of missing speech.
Our neural network models are able to generate new information and they are designed to fill arbitrarily long missing gaps with coherent and plausible signals. 
In addition, we present a Multi-Task Learning (MTL) \cite{caruana1997multitask} strategy where a phone recognition task is learned together with SI. The motivation of the MTL approach lies in previous work, which showed that speech recognition can improve not only speech enhancement \cite{erdogan2015phase} (and vice versa \cite{chen2015speech, pasa2020analysis}), but also speech reconstruction from silent videos \cite{michelsanti_vocoder-based_2020}.

Additional material, which includes samples of the inpainted spectrograms together with the respective audio clips, can be found at the following link: \url{https://dr-pato.github.io/audio-visual-speech-inpainting/}.

\begin{figure*}[!htbp]
    \centering
    \includegraphics[width=.8\textwidth]{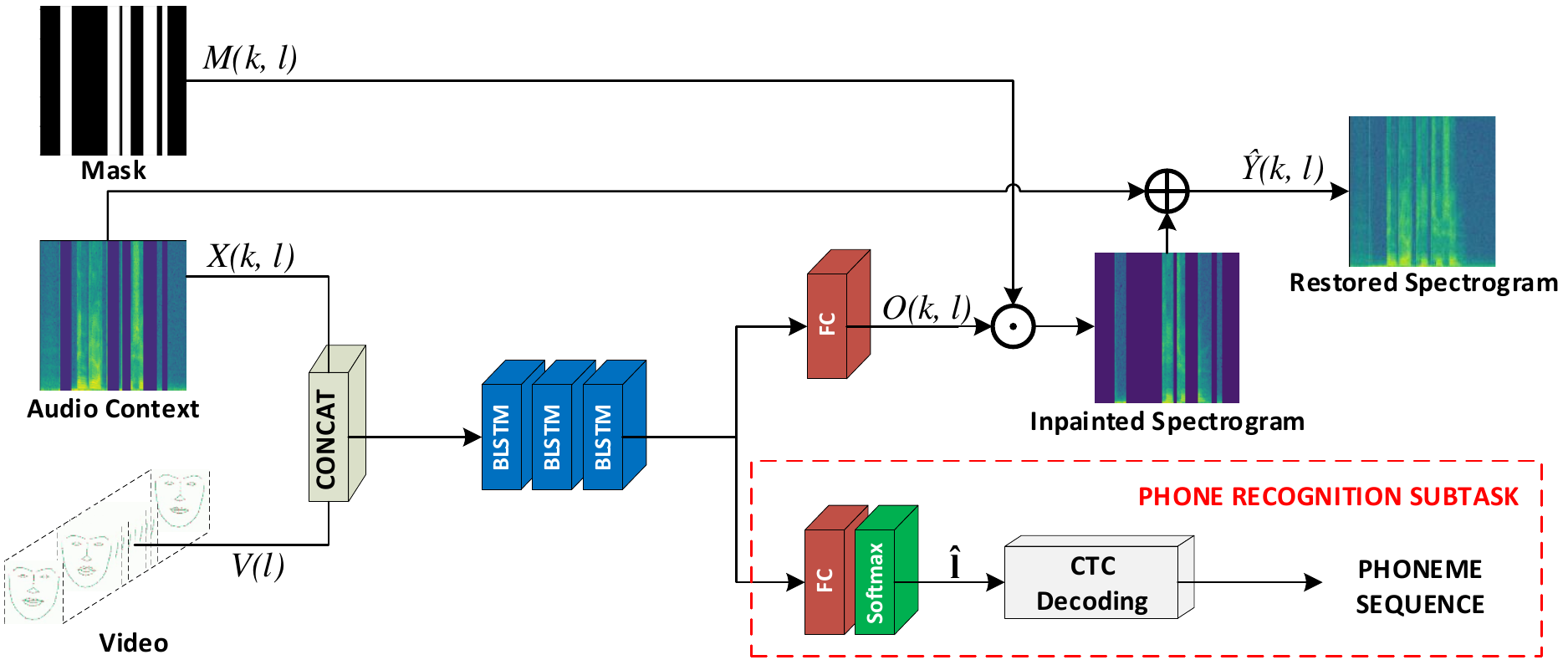}
    \caption{Overall architecture of the audio-visual speech inpainting system. \emph{CONCAT: frame-by-frame concatenation; BLSTM: Bi-directional Long-Short Term Memory unit; FC: Fully Connected layer; CTC: Connectionist Temporal Classification}.}
    \label{fig:architecture}
\end{figure*}

\section{METHODS AND MODEL ARCHITECTURES}
\label{sec:model}
In this section we provide a formulation of the problem and describe the architectures of the models that we propose. As done in previous work \cite{adler_audio_2012}, we assume to know \emph{a priori} the location of reliable and lost data and we use this information in the signal reconstruction stage. In general, the models exploit reliable audio context and visual cues to restore missing gaps in speech signals. As audio and video features, we used log magnitude spectrograms and face landmarks motion vectors, respectively. In a recent work, the specific visual features used here have proven to be effective for audio-visual speech enhancement~\cite{morrone2019face}. 

\vspace{-2mm}
\subsection{Audio-Visual Speech Inpainting Model}
\label{ssec:avmodel}
Let $x[n]$ denote an observed acoustic speech signal, i.e., speech signal with missing parts, with $n$ indicating a discrete-time index. We refer to the log magnitude spectrogram of $x[n]$ as $X(k, l)$, where $k$ and $l$ are a frequency bin index and a time frame index, respectively. The information about the location of missing portions of the signal is encoded in a binary mask $M(k, l)$, which indicates whether a time-frequency tile of the spectrogram of the observed signal is lost, $M(k, l)=1$, or reliable, $M(k, l)=0$. We assume that $X(k, l)=0$ if $M(k, l)=1$. In addition, we denote with $V(l)$ a sequence of visual feature vectors, obtained from the resampled visual frame sequence, since acoustic and visual signals are generally sampled at different rates. 
We define the problem of AV-SI as the task of estimating the log magnitude spectrogram of the ground-truth speech signal, $Y(k,l)$, given $X(k, l)$, $M(k, l)$, and $V(l)$.


In this paper, $Y(k, l)$ is estimated with a deep neural network, indicated as a function, $\mathcal{F}_{av}(\cdot,\cdot,\cdot)$, whose overall architecture is shown in Fig.\ \ref{fig:architecture}. The audio and video features are concatenated frame-by-frame and used as input of stacked Bi-directional Long-Short Term Memory (BLSTM) units that model the sequential nature of the data \cite{graves13}. Then, a Fully Connected (FC) layer is fed with the output of the stacked BLSTM units and outputs the inpainted spectrogram $O(k, l)$. To extract the inpainted spectrogram within the time gaps, $O(k, l)$ is element-wise multiplied with the input mask $M(k, l)$. Finally, the fully restored spectrogram, $\hat{Y}(k, l)$, is obtained by an element-wise sum between the input audio context spectrogram, $X(k, l)$, and the inpainted spectrogram. More formally:
\begin{equation}
\begin{split}
\hat{Y}(k, l) & \triangleq \mathcal{F}_{av}(X(k, l), M(k, l), V(l)) \\
& = O(k, l) \odot M(k, l) + X(k, l)
\end{split}
\end{equation}
where $\odot$ is the element-wise product.
The model is trained to minimize the Mean Squared Error (MSE) loss, $J_{MSE}(\cdot,\cdot)$, between the inpainted spectrogram, $\hat{Y}(k, l)$, and the ground-truth spectrogram, $Y(k, l)$.

\vspace{-5pt}
\subsection{Multi-Task Learning with CTC}
\label{ssec:ctc}
In addition to the plain AV-SI model, we devised a MTL approach, which attempts to perform SI and phone recognition simultaneously. Our MTL training makes use of a Connectionist Temporal Classification (CTC) loss \cite{graves2006connectionist} which is very similar to the one presented in \cite{michelsanti_vocoder-based_2020} for the task of speech synthesis from silent videos. The \emph{phone recognition subtask} block in Fig.\ \ref{fig:architecture} shows the phone recognition module. It is fed with the stacked BLSTM units' output and has a linear FC layer followed by a softmax layer which outputs a CTC probability mass function $\hat{\mathbf{l}} = [\mathbf{p}_1(l), \dots, \mathbf{p}_P(l)]$, with $l \in [1,L]$, where $L$ is the size of the phone dictionary and $P$ is the number of phone labels in the utterance.


The MTL loss function is a weighted sum between the inpainting loss, $J_{MSE}(\cdot,\cdot)$, and the CTC loss, $J_{CTC}(\cdot,\cdot)$:
\begin{equation}
J_{MTL}(Y, \hat{Y}, \mathbf{l}, \mathbf{\hat{l}}) = J_{MSE}(Y, \hat{Y}) + \lambda \cdot J_{CTC}(\textbf{l}, \hat{\textbf{l}}),
\end{equation}
with $\lambda \in \mathbb{R}$, where $\textbf{l}$ is the sequence of ground truth phone labels. 
The phone distribution is used to estimate the best phone sequence. We find the phone transcription applying \emph{beam search} decoding \cite{graves2012sequence} with a beam width of $20$.

\vspace{-.19cm}
\subsection{Audio-only Inpainting Baseline Model}
\label{ssec:amodel}
An audio-only baseline model is obtained by simply removing the video modality from the audio-visual model, leaving the rest unchanged. We consider audio-only models both with and without the MTL approach described in section \ref{ssec:ctc}.

\section{EXPERIMENTAL SETUP}
\label{sec:experiment}
\subsection{Audio-Visual Dataset}
\label{ssec:dataset}
We carried out our experiments on the GRID corpus \cite{cooke_audio-visual_2006}, which consists of audio-visual recordings from $33$ speakers, each of them uttering $1000$ sentences with a fixed syntax. Each recording is $3$ s long with an audio sample rate of $50$~kHz and a video frame rate of $25$ fps. The provided text transcriptions were converted to phone sequences using the standard TIMIT \cite{garofolo1993darpa} phone dictionary, which consists of $61$ phones. However, only $33$ phones are present in the GRID corpus because of its limited vocabulary.

We generated a corrupted version of the dataset where random missing time gaps were introduced in the audio speech signals. Our models are designed to recover multiple variable-length missing gaps. Indeed, for each signal we drew the amount of total lost information from a normal distribution with a mean of $900$ ms and a standard deviation of $300$ ms. The total lost information was uniformly distributed between $1$ to $8$ time gaps and each time gap was randomly placed within the signal. We assured that there were no gaps shorter than $36$ ms and the total duration of the missing gaps was shorter than $2400$ ms. Similarly to \cite{Kegler2020}, the information loss was simulated by applying binary TF masking to the original spectrograms. The generation process was the same for training, validation and test sets.

Our systems were evaluated in a speaker-independent setting, with $25$ speakers (s1-20, s22-25, s28) used for training, $4$ speakers (s26-27, s29, s31) for validation and $4$ speakers (s30, s32-34) for testing. Each set consists of the same number of male and female speakers, except for the training set which contains $13$ males and $12$ females. Furthermore, to evaluate the effect of the gap size, we generated additional versions of the test set, each of them containing a single gap of fixed size ($100$/$200$/$400$/$800$/$1600$ ms).

\vspace{-3mm}
\subsection{Audio and Video Processing}
\label{ssec:processing}
The original waveforms were downsampled to $16$ kHz. A Short-Time Fourier Transform (STFT) was computed using a Fast Fourier Transform (FFT) size of $512$ with Hann window of $384$ samples ($24$ ms) and hop length of $192$ samples ($12$~ms). Then, we computed the logarithm of the STFT magnitude and applied normalization with respect to global mean and standard deviation to get the acoustic input features.

The missing phase was obtained by applying the Local Weighted Sum (LWS) algorithm \cite{le2010fast} to the restored spectrogram. Finally, we computed the inverse STFT to reconstruct the inpainted speech waveform.

We followed the pipeline described in \cite{morrone2019face} to extract the video features, i.e., 68 facial landmarks motion vectors. We upsampled the video features from $25$ to $83.33$ fps to match the frame rate of the audio features.  

\vspace{-4mm}
\subsection{Model and Training Setup}
\label{ssec:setup}
The models in Section \ref{sec:model} consist of $3$ BLSTM layers, each of them with $250$ units. The Adam optimizer \cite{adam} was used to train the systems, setting the initial learning rate to $0.001$. We fed the models with mini-batches of size $8$ and applied early stopping, when the validation loss did not decrease over $5$ epochs. The $\lambda$ weight of the MTL loss, $J_{MTL}(\cdot,\cdot,\cdot,\cdot)$, was set to $0.001$. All the hyperparameters were tuned by using a random search and the best configuration in terms of the MSE loss, $J_{MSE}(\cdot,\cdot)$, on the validation set was used for testing.

\section{RESULTS}
\label{sec:results}
\begin{table}
  \centering
  \resizebox{0.48\textwidth}{!}{%
  \begin{tabular}{cccccccc}
    \toprule
    \multirow{2}{*}{} {A} & {V} & {MTL} && {L1 $\blacktriangledown$} & {PER $\blacktriangledown$} & {STOI $\blacktriangle$} & {PESQ $\blacktriangle$} \\
      \midrule
    \multicolumn{3}{c}{Unprocessed} && $0.838$ & $0.508$ & $0.480$ & $1.634$  \\
    \xmark & & && $0.482$ & $0.228$ & $0.794$ & $2.458$  \\
    \xmark & \xmark & && $0.452$ & $0.151$ & $0.811$ & $2.506$ \\
    \xmark & & \xmark && $0.476$ & $0.214$ & $0.799$ & $2.466$ \\
    \xmark & \xmark & \xmark && $\mathbf{0.445}$ & $\mathbf{0.137}$ & $\mathbf{0.817}$ & $\mathbf{2.525}$ \\
    \bottomrule
  \end{tabular}}
  \caption{Results on the test set. The PER score of uncorrupted speech is $0.069$. \emph{A: audio; V: video; MTL: multi-task learning with CTC.}}
  \label{tab:grid_spk_dep}
\end{table}

\begin{figure*}[t]
    \centering
    \includegraphics[width=1.\textwidth]{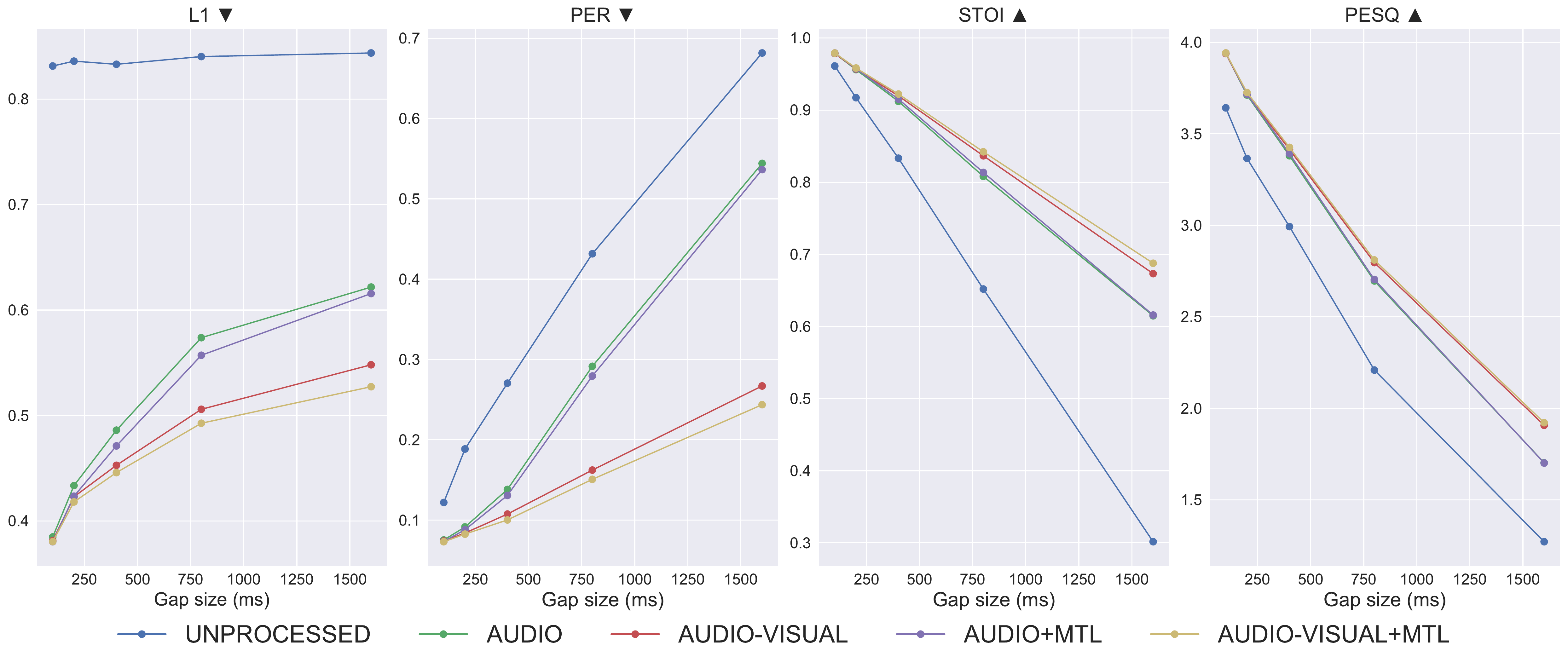}
    \caption{Effect of gap size on speech inpainting performance.}
    \label{fig:gap_charts}
\end{figure*}

\subsection{Evaluation Metrics}
\label{ssec:metrics}
We evaluated the system using L1 loss, Phone Error Rate\footnote{PER was obtained with a phone recognizer trained on uncorrupted data. The phone recognizer consists of $2$ BLSTM layers ($250$ units) followed by a FC and a softmax layers.} (PER), and two perceptual metrics, STOI \cite{taal_algorithm_2011} and PESQ \cite{rix_perceptual_2001}, which provide an estimation of speech intelligibility and speech quality, respectively.
%
While the L1 loss was computed only on the masked parts of the signals, the other three metrics were applied to the entire signals, as it is not possible to perform the evaluation on very short segments. Obviously, PER, STOI, and PESQ show lower sensitivity, when the masked part is small ($< 400$ ms), since a large fraction of the original signal is unchanged in that case. For L1 and PER, lower values are better, while for STOI and PESQ higher values correspond to better performance.

\vspace{-2mm}
\subsection{Discussion}
\label{ssec:discussion}
The evaluation results of the proposed models on the test set are reported in the Table \ref{tab:grid_spk_dep}. On average, the masking process discarded about half of the original speech information, as confirmed by the PER score of unprocessed data. 

Audio-visual models outperform the audio-only counterparts on all metrics, demonstrating that visual features provide complementary information for SI. In particular, the PERs of audio-visual models are lower by a considerable margin, meaning that generated signals are much more intelligible. The improvements in terms of STOI and PESQ are not as large as PER, mainly because the two perceptual metrics are less sensitive to silence than PER. Nonetheless, they are significantly better than the unprocessed data scores confirming the inpainting capability of our models.

The MTL strategy is also beneficial, and results suggest that exploiting phonetic data during the training process is useful to improve the accuracy of SI. However, we observe just a small improvement of the audio-visual MTL model over the plain audio-visual one. This might be explained by the fact that, unlike for the audio-visual system, MTL strategy does not add any additional information at the inference stage.

\subsection{Gap size analysis}
\label{ssec:gap}
Table \ref{tab:grid_spk_dep} reports the average results using multiple variable-length time gaps, not providing information about how the gap size affects the SI capability of our models. For this reason, we generated other test sets, each of them containing samples with a single time gap of fixed length ($100$/$200$/$400$/$800$/$1600$ ms).
Fig.\ \ref{fig:gap_charts} shows the inpainting results for each metric on these test sets. As expected, while for short gaps all models reach similar performance, the difference between audio-only and audio-visual models rapidly increases when missing time gaps get larger. The performance of audio-only models drops significantly with very long gaps ($\geq 800$ ms). Therefore, the audio context does not contain enough information to correctly reconstruct missing audio signals without exploiting vision. In general, audio-only models inpaint long gaps with stationary signals whose energy is concentrated in the low frequencies. On the other hand, audio-visual models are able to generate well-structured spectrograms, demonstrating the benefit that visual features bring to inpaint long gaps. The reader is encouraged to check the difference between audio-only and audio-visual models in the spectrograms provided as additional material (cf. Section \ref{sec:intro}).


Regarding the models trained with the MTL approach, we can notice a good improvement in terms of L1 loss and PER, even if the contribution is not as high as the one provided by the visual modality.


\vspace{-0.3cm}
\section{CONCLUSION}
\label{sec:conclusion}
This work proposed the use of visual information, i.e., face-landmark motion, for speech inpainting. We tested our models on a speaker-independent setting using the GRID dataset and demonstrated that audio-visual models strongly outperformed their audio-only counterparts. In particular, the improvement due to visual modality increased with duration of time gaps.
Finally, we showed that learning a phone recognition task together with the inpainting task led to better results.

\vfill\pagebreak

\bibliographystyle{IEEEbib}
\bibliography{references.bib}

\end{document}